\documentclass[12pt]{article}
\usepackage{amsmath, amssymb,graphicx}

\newwrite\ffile\global\newcount\figno \global\figno=1

\def\writedef#1{}

\input epsf
\def\figin{\epsfcheck\figin}\def\figins{\epsfcheck\figins}
\def\epsfcheck{\ifx\epsfbox\UnDeFiNeD
\message{(NO epsf.tex, FIGURES WILL BE IGNORED)}
\gdef\figin##1{\vskip2in}\gdef\figins##1{\hskip.5in}
\else\message{(FIGURES WILL BE INCLUDED)}%
\gdef\figin##1{##1}\gdef\figins##1{##1}\fi}

\def\figinsert{}
\def\ifig#1#2#3{\xdef#1{fig.~\the\figno}
\writedef{#1\leftbracket fig.\noexpand~\the\figno}%
\figinsert\figin{\centerline{#3}}\medskip\centerline{\vbox{\baselineskip12pt
\advance\hsize by -1truein\center\footnotesize{  Fig.~\the\figno.} #2}}
\bigskip\endinsert\global\advance\figno by1}
\def\endinsert{}

\usepackage{oldgerm}
\usepackage{amssymb}
\begin{document}
\baselineskip 18pt
\renewcommand{\Re}{\mbox{Re}\,}
\renewcommand{\Im}{\mbox{Im}\,}
\renewcommand{\theequation}{\thesection.\arabic{equation}}

\def\N{{\cal N}}

\def\A{{\cal A}}
\def\B{{\cal B}}
\def\O{{\cal O}}
\def\ga{\gamma}
\def\be{\beta}
\def\si{\sigma}
\def\la{{\lambda}}

\def\C{{\cal C}}
\def\O{{\cal O}}
\def\H{{\cal H}}
\def\I{{\cal I}}

\def\al{\alpha}
\def\da{{\dot \alpha}}
\def\db{{\dot \beta}}
\def\dg{{\dot \gamma}}

\def\dphi{\phi \frac{\delta}{\delta \phi} \,} 
\def\dla{\frac{\delta}{\delta \lambda}\, }

\def\tr{{\rm tr}}
\def\e{{\rm e}}
\def\d{{\rm d}}
\def\s{{\rm s}}
\def\pr{\partial}
\def\D{{\cal D}}
\def\L{{\cal L}}

\def\ts{\textstyle}

\renewcommand{\epsilon}{\varepsilon}

\newcommand{\half}{{\textstyle \frac{1}{2}}}
\def\quar{{\textstyle \frac{1}{4}}}

\def\Ga{\Gamma}

\renewcommand{\i}{{\rm i}}

\def\W{{\cal W}}

\def\l{\langle}
\def\r{\rangle}

\def\de{\delta}
\def\si{\sigma}
\def\ga{\gamma}
\def\la{\lambda}
\def\ka{\kappa}

\newcommand{\eps}{\varepsilon}

\def\ts{\textstyle}

\newcommand{\dx}{\!\!{\rm d}^4x\,\,}

\def\E{{\cal E}}
\def\A{{\cal A}}

\newcommand{\gomn}{g^{\mu\nu}}
\newcommand{\gumn}{g_{\mu\nu}}

\def\achtel{{\ts \frac{1}{8}}}


\thispagestyle{empty}
\renewcommand{\thefootnote}{\fnsymbol{footnote}}

{\hfill \parbox{4cm}{
        HU-EP-05-04 \\
ICTP-2005-04 }}

\bigskip \bigskip \bigskip

\begin{center} 

{ \noindent \Large \bf Space-time dependent couplings in 
$\mathcal{N}=1$ SUSY gauge }

\medskip
{\noindent \Large \bf
theories: Anomalies and Central Functions} 

\end{center}

\bigskip\bigskip\bigskip

\bigskip\bigskip \bigskip

\centerline{ \normalsize \bf 
James Babington$^{a,b}$ \footnote[1]{\noindent {\tt
    jbabingt@ictp.trieste.it} } 
  and  Johanna Erdmenger$^a$ \footnote[2]{ \tt jke@physik.hu-berlin.de} }

\bigskip
\bigskip\bigskip

\bigskip

\centerline{ \it $^a$Institut f\"ur Physik, Humboldt-Universit\"at zu Berlin}
\centerline{\it Newtonstra\ss e 15, D-12489 Berlin, Germany}

\bigskip
\centerline{ }

\centerline{ \it $^b$The Abdus Salam 
International Centre for Theoretical Physics}
\centerline{\it Strada Costiera 11, I-34014 Trieste, Italy }

\bigskip\bigskip

\bigskip\bigskip\bigskip\bigskip

\renewcommand{\thefootnote}{\arabic{footnote}}

\centerline{\bf \small Abstract}
\medskip

{\small \noindent We consider $ \mathcal{N}=1 $ supersymmetric gauge theories
  in which the  couplings are allowed to be space-time dependent
  functions. Both the gauge and the superpotential couplings become chiral
  superfields. As has recently been shown, a new topological
  anomaly appears in models with  space-time dependent gauge coupling. Here 
  we show
  how this anomaly may be used to derive the NSVZ $\beta$-function in a
  particular, well-determined renormalisation scheme, both without and
  with chiral matter. Moreover we extend the topological anomaly analysis to
  theories coupled to a classical curved superspace background, and use it to 
  derive an
  all-order expression for the central charge $c$, the coefficient of the Weyl
  tensor squared contribution to the conformal anomaly. We also comment on the
  implications of our results for the central charge $a$ expected to be of
  relevance for a four-dimensional C-theorem.}

\bigskip \bigskip

\newpage

\tableofcontents

\section{Introduction}
\setcounter{equation}{0}

Space-time dependent or `local' couplings $\lambda^i = \lambda^i(x)$ are a
standard tool in quantum field theory. They may be viewed as sources for
composite operators. Well-defined operator 
insertions are obtained by functionally varying the generating
functional with respect to the local
couplings, 
\begin{equation} \label{introins}
\frac{\delta}{\delta \lambda^i (x)} W = \, \langle {\cal O}_i (x)
\rangle
\, .
\end{equation}
The concept of local couplings is particularly appealing 
in supersymmetric theories
where holomorphy is at the origin of non-renormalisation
theorems~\cite{SV1,LK,Seiberg:1993vc,LK2}. 
As we will see below, new results for
supersymmetric gauge theories are obtained by promoting the couplings, both
the gauge couplings as well as the superpotential couplings, to full chiral
and antichiral space-time dependent superfields $\lambda (z) $ and $\bar
\lambda (\bar z) $, respectively, satisfying $\bar D_\da \lambda = 0$ and
$D_\al \bar \lambda = 0$. 

The consequences of allowing for local supercouplings in $\N=1$ theories have
recently been investigated in a series of papers
\cite{Kraus:2001tg}-\cite{Kraus:2001kn} within a perturbative approach. 
Both a component \cite{Kraus:2001id} and a superspace \cite{Kraus:2002nu} 
approach were taken  to study pure $\N=1$
super Yang-Mills theory. The Wess-Zumino model was considered in
\cite{KrausWZ}.

Most importantly, for pure $\N=1$ gauge theory it was shown
in these publications that in the presence of
local couplings there is an additional new anomaly. This anomaly appears in 
the Ward identity for the topological symmetry associated 
with the theta angle and
manifests itself in an anomalous divergence of the topological current. 
At one loop, this anomaly is given by
\cite{Kraus:2002nu} 
\begin{gather} \label{pontanomintro} \left(
\int \! d^6z \, \frac{\delta}{\delta \lambda} - 
\int\! d^6\bar z \, \frac{\delta}{\delta \bar \lambda} \right)
\Gamma 
= \frac{\mathcal{A}_1}{4}\left(\int\! d^6z \,
\frac{1}{\lambda^{\prime}} \, \mathrm{tr}(W^{\alpha}W_{\alpha})-\int\! d^6\bar{z}
\, \frac{1}{
\bar{\lambda}^{\prime} } \, \mathrm{tr}(\bar{W}_{\dot{\alpha}} 
\bar{W}^{\dot{\alpha}}) \right), 
\end{gather}
where $\lambda^{\prime}(z)=\lambda(z)+1/2g^2$ and 
$\bar{\lambda}^{\prime}(\bar{z})=\bar{\lambda}(\bar{z})+1/2g^2$.
$\Gamma$ is the vertex functional of pure $\N=1$ Yang-Mills
theory.\footnote{$\Gamma$ is the vertex functional 
of the BPHZ approach. Within perturbation theory,
by virtue of the so-called `action
  principle', $\Gamma$ is equivalent to the `quantum effective action' 
$\Gamma_{\rm eff}$. This in turn is  
a local function of the fields and of 
the couplings, constructed order by order in perturbation 
theory. In this sense $\Gamma_{\rm eff}$ 
 corresponds to the bare action in the standard perturbative
approach. -  A review of the BPHZ approach
may be found in \cite{BPHZ}.} 
Moreover the one-loop coefficient ${\cal A}_1$ is given by 
${\cal A}_1 = C_2(G)/8 \pi^2$.
The l.h.s. of (\ref{pontanomintro})
 is the symmetry transformation of the vertex 
functional under the topological symmetry. The r.h.s. is the new anomaly which
vanishes in the limit of constant couplings where the integrand becomes the
Pontryagin density. 

The one-loop coefficient of the anomaly in (\ref{pontanomintro}) has been
calculated in \cite{Kraus:2001id} in a component approach. In fact, 
an alternative way of obtaining an equivalent expression for
the lowest component of (\ref{pontanomintro}) is to vary the component 
action with respect to the space-time dependent theta angle. 
This gives rise to the one-loop result  \cite{Kraus:2001id,Bos}
\begin{gather} \label{Jtop1}  \, 
 \frac{\delta}{\delta \tilde \theta (x)} \, \Gamma \, = \, - \, 
 (1 \, + \, {\A}_1 \, 
 \, g^2 )  \left( \quar F^a_{\mu \nu}
  \tilde F^a{}^{\mu \nu}  + \pr^\mu (
\bar \lambda^a \sigma_\mu \lambda^a ) \right) (x)  \, , \qquad \tilde \theta
 = \frac{\theta} {8 \pi^2} \, .
\end{gather}
Here  $\lambda_\alpha{}^a$ 
are the gauginos. (\ref{Jtop1}) implies that the divergence 
of the topological current $J_\mu$, 
defined by
\begin{gather}
J_\mu = \,  \achtel \, \varepsilon_{\mu \sigma \nu \rho} ( A^{\sigma a} \pr^\nu
  A^{\rho a}  + {\ts \frac{1}{3}}
A^{\sigma a} A^{\nu b} A^{\rho c} f_{abc} ) +
\bar \lambda^a \sigma_\mu \lambda^a  \, , \label{Jtop}
\end{gather}
is anomalous. In fact, classically we have $\pr^\mu J_\mu = 
\quar F^{a}_{\mu \nu} \tilde F^a{}^{\mu \nu}  + \pr^\mu (
\bar \lambda^a \sigma_\mu \lambda^a )$ and therefore (\ref{Jtop1}) may
equivalently be written as
\begin{gather}  \, 
 \frac{\delta}{\delta \tilde \theta } \, \Gamma = \, - \, 
(1 + {{\cal A}_1} \, g^2) \,
\pr^\mu J_\mu  \,, 
\qquad {\cal A}_1 = \frac{C_2(G)}{8\pi^2} \, , \label{introcpnts}
\end{gather}
where the anomalous contribution is $\A_1 g^2 \, \pr^\mu J_\mu$. 

In~\cite{Kraus:2002nu}, the anomaly in (\ref{pontanomintro})
is used to obtain  a general scheme independent result for the
gauge $\beta$ function to all orders. 
The main point in this derivation is to add an appropriate counterterm
to the action which shifts the anomaly from the topological Ward identity
(\ref{pontanomintro})  to the Callan-Symanzik equation. 
There is an interesting analogy between this anomaly shift and the approach 
of~\cite{Arkani-Hamed:1997mj},  where  anomalous transformations of the path
integral functional measure are considered. 
- Note in particular that the one-loop coefficient 
${\cal A}_1 = C_2(G)/8 \pi^2$ is the coefficient which appears in 
the denominator of the NSVZ $\beta$-function~\cite{Novikov:ic}.

A further use of local couplings, so far used for non-supersymmetric field
theories or for SUSY theories in components, is that they allow for an elegant
formulation of {\it local} renormalisation group (RG) or Callan-Symanzik (CS)
equations. These determine how conformal symmetry is broken in a quantised
field theory. This is in contrast to the standard (or {\it global}) RG
equations, which express the breakdown of scale invariance upon quantisation. 
For the formulation of local RG equations it is necessary to couple the
quantum field theory to a classical curved space background, with the metric
acting as source for the energy-momentum
tensor~\cite{Osborn:gm,Kraus:1992ru} (see also \cite{Jack}). 
The significance of this approach is that it
may be of relevance for a proof of an analogue of 
the Zamolodchikov C-theorem~\cite{Zamolodchikov} in
more than two dimensions. In particular in~\cite{Osborn:gm} an alternative
derivation of the two-dimensional C-theorem was given using local
couplings. It is essential for this analysis that in the presence of local
couplings, there are new conformal anomalies in addition to the familiar
anomalies involving the curvature of the background metric. 
These new conformal anomalies involve derivatives of the
couplings. Furthermore in \cite{Osborn:gm} within the local coupling approach,
the flow of a candidate {\it a}-function (related to the coefficient of the
Euler anomaly) for a possible four-dimensional C-theorem~\cite{Cardy:cw} is
related to a quadratic form, which however so far has not been shown to be positive definite. 

A related approach to supersymmetric theories was used in~\cite{Osborn:2003vk}
and in particular in~\cite{FreedmanOsborn} where a four-loop expression for
the candidate {\it a}-function was given and shown to coincide with
non-perturbative results for $a_{\rm UV} - a_{\rm IR}$ found 
in~\cite{Anselmi:1997am}. These non-perturbative results were obtained using
't Hooft anomaly matching and by performing explicit one-loop computations of
three-point correlators involving the R current, the energy-momentum
tensor and a particular anomaly-free current.

A different approach to a possible four-dimensional C-theorem for
supersymmetric theories is the principle called 
`$a$ maximization' which has recently been proposed 
and investigated 
in 
\cite{Intriligator:2003jj}-\cite{Kutasov:2004xu}. The local coupling approach taken here is
complementary to $a$ maximization, though relations between the two approaches
exist \cite{Barnes:2004jj}.

The purpose of this paper is twofold. First we consider 
$\mathcal{N}=1$ SUSY gauge
theories with local chiral supercouplings both without and with chiral matter. 
We show how the NSVZ $\beta$-function may easily be derived from the 
topological anomaly present for local couplings. 
This $\beta$-function is naturally
associated to a particular renormalisation scheme which we describe in
detail. 

Secondly we consider $\N=1$ SUSY gauge theories with local couplings which in
addition are coupled to a classical supergravity background in
superspace. The aim of this analysis is to find new results for the
coefficients of the gravitational anomalies. -   
For the superspace formulation of
local CS equations, the supercurrent associated to
superconformal transformations 
has to be coupled to the appropriate superspace supergravity 
field~\cite{Buchbinder:qv,Gates:nr}.  An analysis of local CS equations for 
supersymmetric theories with constant coupling has been given 
in~\cite{Erdmenger:1998tu,Erdmenger:1998xv,Erdmenger:1999uw}. 
Here we extend this approach to the local coupling case. 
The resulting superconformal Ward identities may  be viewed as a
generalisation of previous results for the anomalous divergence of the
supercurrent~\cite{Clark:1980dw,Shifman:1986zi,Kogan:1995mr,Leigh:1995ep} 
to the off-shell
case with curved superspace background. With this approach we are able to give
an all-order derivation of an expression for the central charge $c$, the
coefficient of the Weyl tensor squared contribution to the conformal anomaly,
in terms of the beta and gamma functions of the theory, of the form
\begin{gather} \label{cc}
c = c_1 + \frac{1}{24}\left( N_V \frac{\beta_g}{g} - 
{\gamma_i} {N_\chi{}^i} \right) \, , \;\; c_1 = {
\frac{1}{24}}\, 
(3 N_V + 
N_\chi) \, , \qquad  N_\chi =\sum\limits_{i} N_\chi{}^i \, .
\end{gather}
Here $N_\chi{}^i$ denotes the number of chiral fields with anomalous
dimension $\gamma_i$ (We take the anomalous dimension matrix to be
diagonal). $c_1$ is the one-loop result for the central charge.
The expression (\ref{cc}) was
first presented in~\cite{Anselmi:1996dd}, and is
based on the two-loop calculations 
of~\cite{Jack:wd}. Our all-order 
derivation presented here relies on the fact that on curved
superspace, the topological Ward identity has an additional one-loop
anomaly of the form 
\begin{gather} \label{pontanomintro2} \left(
\int\! d^6z \,\frac{\delta}{\delta \lambda} - 
\int\! d^6\bar z \,\frac{\delta}{\delta \bar \lambda} \right)
\Gamma^\prime 
= \frac{\mathcal{C}_{1}}{4}\left(\int\! d^6z \,
\frac{1}{\lambda^{\prime}} \mathrm{tr}(W^{\alpha\beta\gamma}W_{\alpha\beta \gamma})
-\int\! d^6\bar{z}
\, \frac{1}{
\bar{\lambda}^{\prime} } \mathrm{tr}(\bar{W}_{\dot{\alpha}\db\dg} 
\bar{W}^{\dot{\alpha}\db\dg}) \right),
\end{gather}
where $W_{\al \beta\gamma}$ is the superspace Weyl density. $\Gamma'$ is the
vertex functional in which a suitable local counterterm has been added to
$\Gamma$ in
(\ref{pontanomintro}) such as to cancel the r.h.s.~of (\ref{pontanomintro}). 
Note again that
the anomaly (\ref{pontanomintro2}) vanishes in the constant coupling limit. 
We calculate the coefficient 
${\mathcal C}_1$ to one loop in the component decomposition of
\cite{Anselmi:1997am}, and
show how it gives rise to the desired result for $c$ in the same
renormalisation scheme as used for 
the derivation of the NSVZ $\beta$ function. 
Moreover we extend our result to theories with matter by
considering an off-shell version of the Konishi anomaly on curved
space background. 
The essential point of our derivation is again - as in \cite{Kraus:2002nu} - 
that with a suitable local 
counterterm, the anomaly in (\ref{pontanomintro2}) may be
shifted from the topological Ward identity to the superconformal Ward identity
and thus to the Callan-Symanzik equation. 

The coefficient of the Euler central charge $a$ is inherently more difficult
to derive in the approach presented here, 
since it contains terms non-linear in the
anomalous dimension, of the form $\tr(\gamma \gamma)$ and $\tr(\gamma \gamma
\gamma)$. We expect to return to a derivation of $a$
in the future. At least already
at the present stage we are able to show that there cannot be any
terms of the form  ${\rm tr} (\gamma) $ contributing to $a$. 
Moreover our results are consistent with the expected factor of the
$\beta(g)/g$ contribution to $a$. 

The outline of this paper is as follows. 
We begin in section~\ref{internal} with a brief summary of the results of \cite{Kraus:2002nu} 
relevant for our analysis, the
implications of local chiral couplings in supersymmetric Yang-Mills theory
and the associated internal anomalies. Then we identify the constraints
leading to the particular renormalisation scheme which gives rise to the
NSVZ $\beta$-function. Moreover we extend the analysis to
gauge theories with matter.
In section~\ref{external} we find expressions for the central
charge $c$ by considering the external
anomalies. Firstly, the gauge field contribution to the new topological
anomaly is calculated and is shown to give rise to
the $\beta (g)/g$ contribution to $c$. Next we include the matter
contributions by 
considering the Konishi anomaly in an off-shell approach for a curved
superspace background.
Finally in section~\ref{conclusion} we
conclude with comments on the implications for the Euler central charge $a$ 
and an outlook on future directions.
The appendix contains the necessary one-loop triangle diagram computations.

\section{Local Chiral Couplings - Internal anomalies}\label{internal}
\setcounter{equation}{0}

We begin by giving a summary of the results in~\cite{Kraus:2002nu,Kraus:2002se} which show that in a pure gauge theory, a new
topological anomaly appears when the couplings are allowed to be space-time
dependent. We then show how the NSVZ beta function arises in this approach for
a particular renormalisation scheme.

\subsection{Topological anomaly in pure gauge theory}

The starting point is pure $ \mathcal{N}=1 $ SYM with gauge group $G$, whose
classical action is, in the conventions of~\cite{Kraus:2002nu},
\begin{equation}
S_{{\rm constant \, coupling}}[V]=\, - \frac{1}{4g^2}  \int\! d^6z \,
\mathrm{tr}(W^{\alpha}W_{\alpha}) - \frac{1}{4g^2}\int\! d^6\bar{z} \, 
\mathrm{tr}(\bar{W}_{\dot{\alpha}} \bar{W}^{\dot{\alpha}}),
\end{equation}
with
\begin{gather} \label{ws}
W_\alpha=\,\frac{1}{8}\bar{D}^2(e^{-2V}D_{\alpha}e^{2V}), \qquad
\bar{W}_{\dot \alpha}=-\frac{1}{8}D^2(e^{2V}\bar{D}_{\dot{\alpha}}e^{-2V}).
\end{gather}
Next  the gauge coupling  $ g $ is promoted to
 a local chiral and antichiral superfield, $ \lambda(z) $ and $
 \bar{\lambda}(\bar{z}) $ respectively,  such that the action becomes
\begin{equation}\label{action1}
{\cal S} [V]=\, - 
\frac{1}{2}\int \!d^6z\, \lambda(z) \mathrm{tr}(W^{\alpha}W_{\alpha})
- \frac{1}{2}\int \! d^6\bar{z}\, 
\bar{\lambda}(\bar{z}) \mathrm{tr}(\bar{W}_{\dot{\alpha}} \bar{W}^{\dot{\alpha}}).
\end{equation}
As discussed in \cite{Kraus:2001id},  
the vector superfield $ V $ may \emph{not} be fixed to the Wess-Zumino gauge if 
manifest supersymmetry is to be preserved in the presence of local couplings. In the case where the 
Wess-Zumino gauge is fixed, the supersymmetry algebra closes 
only up to gauge transformations and 
hence is  not linearly realised. 

In order to be able to make recourse to perturbation theory, there has to be a
well-defined free field limit. For the action (\ref{action1}),
there are two ways to proceed:
If  the lowest components of $\lambda(z)$ and $\bar{\lambda}(\bar{z})$ are
taken to coincide with the local coupling and local theta angle by virtue of
\begin{gather} \label{coupling2}
\lambda(z)|_{\theta}+\bar{\lambda}(\bar{z})|_{\bar{\theta}}
=\frac{1}{2g^2(x)}, \quad
\lambda(z)|_{\theta}-\bar{\lambda}(\bar{z})|_{\bar{\theta}}
=\, - \frac{i}{16\pi^2}\Theta(x)
\, ,
\end{gather}
then a rescaling of $V$ by
\begin{gather}
V\rightarrow (\lambda+\bar{\lambda})^{-1/2}V \, ,
\end{gather}
leads to a well-defined free field limit of the action (\ref{action1}). 
It would be very interesting to study the renormalisation behaviour of the
action with this normalization. 

Here, however, we follow another approach for
calculational simplicity, first used in \cite{Kraus:2002nu}.
In this approach a well-defined free field limit is obtained by shifting
the lowest components of the superfields by a constant such that
\begin{gather} \label{coupling}
\lambda^{\prime}(z) = \lambda(z)+\frac{1}{2g^2} \, , \qquad
\bar{\lambda}^{\prime}(\bar{z}) = \bar{\lambda}(\bar{z})+\frac{1}{2g^2}.
\end{gather}
The new action with these couplings reads
\begin{gather} \label{gauge}
S[V]=\, - 
\frac{1}{2}\int\! d^6z\,\lambda^{\prime}(z) \mathrm{tr}(W^{\alpha}W_{\alpha})
- \frac{1}{2}\int\! d^6\bar{z}\,\bar{\lambda}^{\prime}(\bar{z})
\mathrm{tr}(\bar{W}_{\dot{\alpha}} \bar{W}^{\dot{\alpha}}) \, .
\end{gather}
From this classical action a perturbative expansion for the associated
quantum theory is obtained as follows.
Varying the vertex functional corresponding to (\ref{gauge})
with respect to $\lambda$ or $\bar \lambda$ gives rise to well-defined
operator insertions, which in turn correspond to the vertices of the
perturbation expansion. 
Moreover the action (\ref{gauge})
has a well-defined free field limit, which is obtained by
setting $ \lambda=0, \bar{\lambda}=0 $.
This allows for an unambiguous definition of the free propagator 
but with an additional factor of $g^2$ in its definition as compared to the
standard constant coupling perturbative approach. 
Kraus and collaborators show that the perturbative expansion obtained
in this way - vertices from functional differentiation and modified
propagators - satisfies
\begin{equation} \label{ng}
N_g=2(N_{\lambda}+N_{\bar{\lambda}})+2(l-1)+N_{V},
\end{equation}
where $N_{g}$ denotes the power of the constant coupling $g$ 
in a Feynman graph, 
$N_{\lambda}$ and $N_{\bar{\lambda}}$ count the
operator insertions obtained by varying with respect to $\lambda$ or
$\bar  \lambda$, respectively, and $N_{V}$ the number of external
superfield legs. $l$ is the number of loops. 
The relation (\ref{ng}) ensures that
the loop expansion is a power series.

For the action (\ref{gauge}) there are
two important classical Ward identities which become
anomalous upon  quantisation. These arise due to a new global symmetry
associated with the local couplings. The relevant
 symmetry transformation consists of shifting the 
local couplings by a complex constant $\omega$,
\begin{gather} \label{coupling3}
\lambda(z)\rightarrow \lambda(z)+\omega, \qquad 
\bar{\lambda}(\bar{z})\rightarrow\bar{\lambda}(\bar{z})+\bar{\omega}.
\end{gather} 
If we define the following operators
\begin{equation} \label{plusminus}
 \Delta^{\pm}\equiv \int\!
 d^6z\,\frac{\delta}{\delta\lambda(z)}\pm\int\! d^6\bar{z}\,
\frac{\delta}{\delta\bar{\lambda}(\bar{z})},
\end{equation} 
then it is easily verified that the classical action $ S[V] $ 
satisfies the {\it shift
  equation}, induced by the real part of $\omega$,
\begin{equation} \label{shift}
\Delta^{+}S=-g^3\partial_gS \, .
\end{equation} 
This identity is crucial for establishing the equivalence between the
perturbation expansion described above and the standard one.

In addition, there is also the {\it Pontryagin identity} arising from
the imaginary part
of $\omega$,
\begin{equation}
\Delta^{-}S=\, - \frac{1}{2}\int\! d^6z \,
\mathrm{tr}(W^{\alpha}W_{\alpha}) + \frac{1}{2}\int\!
d^6\bar{z}\, \mathrm{tr}(\bar{W}_{\dot{\alpha}} \bar{W}^{\dot{\alpha}})=0 .
\end{equation}
The r.h.s. vanishes on flat space since it is an integral over a topological density.

The crucial result of \cite{Kraus:2002nu} is that in the quantised theory, the
Pontryagin identity becomes anomalous. At one loop, the anomaly is given by
\begin{equation} \label{pontanom}
\Delta^{-}\Gamma= \frac{\mathcal{A}_{1}}{4}\left(\int\! d^6z \,
\frac{1}{\lambda^{\prime} (z)} \mathrm{tr}(W^{\alpha}W_{\alpha})-\int\! d^6\bar{z}
\, \frac{1}{
\bar{\lambda}^{\prime}(\bar{z})} \mathrm{tr}(\bar{W}_{\dot{\alpha}}
\bar{W}^{\dot{\alpha}}) \right) \, ,
\end{equation}
with
$\Gamma$ the BPHZ  vertex functional.
(Capital calligraphic letters denote anomaly coefficients and the number 1
denotes 1-loop.)  
The anomaly coefficient $ \mathcal{A}_1 $ was calculated in~\cite{Kraus:2001id} 
and is given by
\begin{equation}
\mathcal{A}_1=\frac{1}{8\pi^{2}}C_2(G),
\end{equation}
where the group theory factor is $ (T^AT^A)_{R}=C(R) {\bf 1} $ for a representation $R$ of the gauge group, and $R=G$ is the adjoint representation. For $G= SU(N_c) $ we have
\begin{equation}
\mathcal{A}_1=\frac{N_c}{8\pi^{2}}.
\end{equation} 
It is essential to note that the local couplings enter
the anomaly in (\ref{pontanom}) in the form
$1/\lambda'$, $1/{\bar \lambda}'$. 
This is consistent with the fact that when taking the constant
coupling limit, the integrand of (\ref{pontanom}) reduces
to a component form that contains a factor of $g^2$ as in (\ref{introcpnts})
(see  \cite{Kraus:2001id}).

\subsection{Shifting the anomaly}

We have seen that the Pontryagin equation has an anomaly given by
(\ref{pontanom}). However by using the freedom of adding local 
counterterms, 
$ \Gamma^{ct}(V) $, to the quantum action, it is possible to move
the anomaly 
to the shift equation (\ref{shift}). This is
analogous to the situation for the chiral anomaly, where
one has the freedom to shift the anomaly between the axial  and 
vector Ward identities. The specific counterterm chosen
in~\cite{Kraus:2002nu}, which shifts the anomaly from (\ref{pontanom}) to
(\ref{shift}),  is 
\begin{gather}
\Gamma^{ct}(V)= \, - \frac{\mathcal{A}_1}{4}\left(\int\!
d^6z\, [\ln(2\lambda^{\prime}(z))+\ln g^2] \mathrm{tr}(W^{\alpha}W_{\alpha})+\int\!
d^6\bar{z}\, [\ln(2\bar{\lambda}^{\prime}(\bar{z}))+\ln g^2]
\mathrm{tr}(\bar{W}_{\dot{\alpha}} \bar{W}^{\dot{\alpha}}) \right). \label{counterterm1}
\end{gather}
The $ \ln g^2 $ terms in this expression are
necessary to ensure that this counterterm vanishes in the constant coupling limit
$ \lambda=0, \bar{\lambda}=0 $, such that the constant coupling perturbation
expansion remains a power series.
The action
\begin{gather} \label{Gammaprime}
\Gamma^{\prime}=\Gamma+\Gamma^{ct}(V)
\end{gather} 
satisfies the Pontryagin identity
\begin{gather}\label{ponteq}
\Delta^{-}\Gamma^{\prime}=0 \, .
\end{gather} 
However the shift equation now becomes anomalous,
\begin{equation} \label{qshift0}
\Delta^{+}\Gamma^{\prime}=-g^3\partial_{g}
\Gamma^{\prime} - \frac{\mathcal{A}_1g^2}{2}\left(\int\!
  d^6z\,\mathrm{tr}(W^{\alpha}W_{\alpha})
+\int\! d^6\bar{z} \, \mathrm{tr}(\bar{W}_{\dot{\alpha}} \bar{W}^{\dot{\alpha}}) \right).
\end{equation} 
The anomaly term is just the classical SYM action being acted upon by the
operator $ \Delta^{+} $, such that at one loop, (\ref{qshift0}) may be written as
\begin{gather}\label{qshift}
\Delta^{+}\Gamma^{\prime}=-g^3\partial_{g}\Gamma^{\prime}+\mathcal{A}_1g^2\Delta^{+}S.
\end{gather}
This  result is scheme independent. 

We now proceed by chosing a particular scheme, which we show to
coincide with the NSVZ scheme.
A particular scheme is obtained by assuming that (\ref{ponteq}) and 
(\ref{qshift})
are valid to {\it all}  orders in perturbation theory, which
corresponds to replacing (\ref{qshift}) by the full quantum equation
\begin{gather} \label{qshift2}
\Delta^{+}\Gamma^{\prime}=-g^3\partial_{g}
\Gamma^{\prime}+\mathcal{A}_1g^2\Delta^{+}\Gamma^{\prime}.
\end{gather}
This scheme choice requires in particular that
the Pontryagin anomaly we discussed above is exhausted at one-loop in this
particular scheme.\footnote{It appears feasible to prove one-loop
exactness of the Pontryagin anomaly by an argument based on the
results of \cite{Bos} together with supersymmetry. However finite
renormalisations at higher order are always possible.}

It is straightforward to see that the scheme
defined by passing from (\ref{qshift}) to (\ref{qshift2}) is indeed the NSVZ
scheme in which the beta function takes the form 
derived by NSVZ in~\cite{Novikov:ic}. 
Indeed, a rearrangement of~(\ref{qshift2}) gives
\begin{gather} \label{shifta}
\Delta^{+}\Gamma^{\prime}=-\frac{g^3}{1-\mathcal{A}_1g^2}
\partial_{g}\Gamma^{\prime}.
\end{gather}
This has already a form reminiscent of the NSVZ $ \beta
$-function. For an exact identification we now
establish the connection to scale transformations
of the vertex functional as expressed by a renormalisation group
flow. 

As discussed in \cite{Kraus:2002nu}, when (\ref{ponteq}) holds to all
orders, the local coupling $\lambda$ (or $\bar \lambda$) is not
renormalised beyond one loop and remains holomorphic (or antiholomophic)
in the quantised theory. This is due to the consistency condition 
\begin{gather} \label{consistency1}
\left[ \,  \Delta^- \, , \; \mu \frac{\partial}{\partial \mu}  \; \right] 
\Gamma^\prime = 0 \, ,
\end{gather}
with $\Delta^-$ defined in (\ref{plusminus}) and $[ \, , \,  ]$ the
commutator.  
Therefore the Callan-Symanzik equation reads
\begin{gather} \label{CS1}
\left( \mu \frac{\pr}{\pr \mu} \, + \, {\cal B}_1 \Delta^+ \right)
\, \Gamma^\prime \, = 0 \, , \qquad {\cal B}_1 = \frac{3}{16 \pi^2}
C_2(G) \, ,
\end{gather}
with $\Delta^+$ defined in (\ref{plusminus}) and ${\cal B}_1$ the
standard one-loop coefficient of the gauge beta function.
Inserting (\ref{shifta}) into (\ref{CS1}) then gives rise to
to the Callan-Symanzik equation
\begin{gather} \label{CS}
\left(\mu\frac{\partial}{\partial \mu}+\beta(g)\partial_{g}
\right)\Gamma^{\prime}=0 \, ,
\end{gather}
with
\begin{equation}
\beta(g) = -\frac{\mathcal{B}_1g^3}{1-\mathcal{A}_1g^2}=-\frac{g^3}{16\pi^{2}}\left(\frac{3C_2(G)}{1-C_2(G)g^2/(8\pi^2)}\right),
\end{equation} 
which is precisely the NSVZ $\beta$-function.

A further important point is that if (\ref{ponteq}) is valid to all
orders in perturbation theory,  then a well-defined
local operator insertion is obtained by virtue of
\begin{gather} \label{localins}
\frac{\delta}{\delta \lambda(z)} \, \Gamma^\prime \, = \, - \,
\frac{1}{2} \, 
{\rm tr} \, W^\al W_\al (z)  \, .
\end{gather} 
In standard notation, local insertions of composite operators are often
denoted by square brackets, ie.~$[W^\al W_\al]$, but we omit this here
and in the remainder of this paper to simplify the notation.  

The Callan-Symanzik equation (\ref{CS1}) may be interpreted as an
anomalous Ward identity for scale transformations. 
Scale transformation are a subgroup of superconformal transformations,
and together with the local equation (\ref{localins}), (\ref{CS1})
implies that the superconformal Ward identity consistent with
(\ref{CS1}) is given by
\begin{gather} \label{oneloop}
\bar D^\da T_{\al \da}\,   = \, D_\al T \, , \qquad T = - \frac{1}{6} {\cal
B}_1 \,  \tr \, W^\beta W_\beta  \, .
\end{gather}
Here $T_{\al \da}$ is the supercurrent from which all currents of the
superconformal group may be obtained. Note that for the theory with
local couplings considered here,  the supercurrent has contributions
which involve derivatives of the couplings. Moreover the
superconformal anomaly $T$ is one-loop.

\subsection{Curved superspace background}

We proceed by deriving a local version of the Callan-Symanzik equation
(\ref{CS}). 
For this purpose we couple the quantised local coupling theory to a
classical curved superspace background. As in \cite{Buchbinder:qv,Gates:nr},  
the curved superspace involves a chiral compensator $\phi$ and a real
Weyl invariant superfield $H^{\al \da}$, such that local quantum insertions of
the supercurrent $T_{\al \da}$ and of the superconformal anomaly $T$
are given by
\begin{gather}
\label{scano}
\bar {\cal D}^\da T_{\al \da} \, = \, {\cal D}_\al  T \; , 
\qquad T_{\al \da} = 
\frac{\delta}{\delta H^{\al \da}} \Gamma^\prime \, , \quad T \, = \, \frac{1}{3}
\, \phi
\frac{\delta}{\delta \phi} \Gamma^\prime \, .
\end{gather}
It is important to note that in the case of constant couplings, $T$ as
given
by (\ref{scano}) has higher-order quantum corrections. It is possible to
redefine the supercurrent $T_{\al \da}$ such that $T$ is one-loop, but then
the coupling to curved superspace as given by (\ref{scano}) 
is inconsistent (see
\cite{supercurrentPS,supercurrentGZ,Piguet:1986ug,Gates:nr}).

However for the local coupling theory considered here, the curved superspace
background given by (\ref{scano}) is consistent with $T$ being
one-loop, as we now show. This is possible essentially since the
supercurrent is modified by additional terms involving derivatives of
the couplings.

When  
coupled to a curved superspace background as in
\cite{Buchbinder:qv,Gates:nr},
the classical action (\ref{gauge}) becomes
\begin{gather} \label{gaugecurved}
S[V]=\, -  
\frac{1}{2}\int\! d^6z\, \phi^3 \, 
\lambda^{\prime}(z) \mathrm{tr}(W^{\alpha}W_{\alpha})
- \frac{1}{2}\int\! d^6\bar{z} \, \bar \phi^3 \, \bar{\lambda}^{\prime}(\bar{z})
\mathrm{tr}(\bar{W}_{\dot{\alpha}} \bar{W}^{\dot{\alpha}}) \, .
\end{gather}
The topological anomaly discussed in section 2.1 is present also on the curved
superspace background. A quantum action $\Gamma'$ satisfying the topological
Ward identity $\Delta^- \Gamma^\prime \, = \, 0$ as in (\ref{ponteq}) 
is obtained from the quantum action $\Gamma$ corresponding to the classical
action (\ref{gaugecurved}) by adding a suitable local counterterm,
$ \Gamma^\prime \, = \, \Gamma \, + \, \Gamma^{ct}(V)$ given by
\begin{align} \label{counterterm2}
\Gamma^{ct}(V) = -& \frac{\mathcal{A}_1}{4} \int\!
d^6z \,\phi^3 [\ln(2\lambda^{\prime}(z))+\ln g^2]
 \mathrm{tr}(W^{\alpha}W_{\alpha}) \nonumber \\
 -& \frac{\mathcal{A}_1}{4} \int\! d^6\bar{z}\, \bar \phi^3
 [\ln(2\bar{\lambda}^{\prime}
(\bar{z}))+\ln g^2]
\mathrm{tr}(\bar{W}_{\dot{\alpha}} \bar{W}^{\dot{\alpha}}) \, , 
\end{align}
which is the curved superspace analogue of (\ref{counterterm1}). 
This counterterm is Weyl invariant. 

The Ward identity  $\Delta^- \Gamma^\prime \, = \, 0 $
is crucial for our construction. It implies that 
a local operator insertion is obtained by virtue of
\begin{gather}
\frac{\delta}{\delta \lambda} \, \Gamma^\prime \, =  \, \phi^3 \,
\tr \, W^\al W_\al  \, .
\end{gather} 
Then
the consistency condition
\begin{gather} \label{consistency2}
\left[ \, \phi \frac{\delta}{\delta \phi}  \, , \, \frac{\delta}{\delta
    \lambda} \, \right] \, \Gamma^\prime \, = \, 0
\end{gather}
implies that
\begin{gather} \label{TTT}
T \, = \, \frac{1}{3} \,  \phi \frac{\delta}{\delta \phi} \, \Gamma^\prime 
\end{gather}
is one-loop. 
Note that $T_{\al \da}$,  the
insertion of the supercurrent given by (\ref{scano}), 
contains derivatives of the local couplings and
is therefore different from the supercurrent in the standard constant
coupling theory. 

By dimensional analysis, the
quantum vertex functional $\Gamma^\prime$  satisfies
the scale relation
\begin{gather} \label{cseqn}
\mu\frac{\partial}{\partial\mu}\Gamma^{\prime}\, + \,   \left(\int\!
 d^6z\, \phi\frac{\delta}{\delta\phi}\, + \int\! 
d^6\bar{z}\, \bar{\phi}\frac{\delta}{\delta\bar{\phi}}\right)\Gamma^{\prime}
\, = \, 0 \, .
\end{gather}
The consistency condition (\ref{consistency2}) implies that 
the Callan-Symanzik equation of the flat space case (\ref{CS1}) 
is also valid for the theory coupled to curved superspace,
\begin{gather} \label{CS3}
\left( \mu \frac{\pr}{\pr \mu} \, + \, {\cal B}_1 \Delta^+ \right)
\, \Gamma^\prime \, = 0 \, , \qquad {\cal B}_1 = \frac{3}{16 \pi^2}
C_2(G) \, .
\end{gather}
A local Callan-Symanzik equation consistent with
(\ref{CS3}) and (\ref{cseqn}) is given by
\begin{gather} \label{susylocalRG}
\phi \frac{\delta}{\delta \phi}  \Gamma^\prime \, = \, 
{\cal B}_1  \,  
\frac{\delta}{\delta \lambda }  \Gamma^\prime 
\, .
\end{gather}
The superconformal anomaly reads
\begin{gather} \label{wti}
\bar {\cal D}^\da T_{\al \da} \, = \, {\cal D}_\al  T \; , \qquad
T \, = \, - \frac{{\cal B}_1}{6} \,  
 \phi^3 \, {\rm tr} W^\al W_\al \, .
\end{gather} 
This is the curved superspace analogue of (\ref{oneloop}).


\subsection{Theories with matter}
Lets us now consider the action of a gauge theory with matter,
\begin{align}
S = S_{\rm gauge} + S_{\rm matter} \, , \label{ac}
\end{align}
where the gauge part of the action is as in (\ref{gauge}) above, 
and the matter part contains
$n$ chiral fields $\Phi^i$ transforming
in the representations $R_i$ of the gauge group $G$. 
On a curved space background, the matter action is given by 
(see~\cite{Buchbinder:qv,Gates:nr} for 
superspace notation)
\begin{equation}\label{matteract}
S_{\rm matter}
= \, \frac{1}{4} \, \int \! d^8  z \, \tilde E\bar{\Phi}_i e^{2V}  \Phi^i \, .
\end{equation} 
$\tilde E$ is the appropriate curved superspace integration measure and
$i$ is the flavour index. We suppress colour indices for notational
simplicity. We first restrict to the case where there is no
superpotential, and include it in a second step below.

The new vertex functional corresponding to (\ref{ac}) 
still satisfies the Pontryagin
equation~(\ref{ponteq}) with the same anomaly as before. Therefore
the shift equation~(\ref{qshift2}) is preserved. This implies in particular
that the denominator for the NSVZ $\beta$-function is unchanged as expected.

We now generalise the local Callan-Symanzik equation 
(\ref{susylocalRG}) to the action (\ref{ac}) which includes chiral
matter. Our starting point is the result 
of~\cite{Shifman:1986zi,Kogan:1995mr,Leigh:1995ep}, 
based on an earlier result in~\cite{Clark:1980dw},
according to which on flat space, 
the supercurrent anomaly for the action (\ref{ac})
 may be written as
\begin{gather} \label{LS}
\bar D^{\dot \alpha} T_{\al \da} = \, - \frac{1} {3} \, D_\al \left(
  \frac{{\mathcal B}_1{}'}{2} {\rm tr} (W^\beta W_\beta) 
+ \, \frac{1}{4} \bar D^2
  \sum\limits_{i=1}^{n} \gamma_i 
  \bar \Phi_i e^{2V} \Phi^i \right) \, ,
\end{gather}
where the coefficient of the gauge anomaly is one-loop and $\gamma_i$ is the
anomalous dimension\footnote{As in~\cite{Leigh:1995ep}, we assume that 
the anomalous dimension
matrix is diagonal, 
$\gamma^i{}_j = \gamma_{(i)} \delta^i{}_j$. Note also that the
anomalous dimension of $\Phi^i$ used here is half the value of the mass
anomalous dimension used 
in~\cite{Kogan:1995mr,Leigh:1995ep,Anselmi:1997am}. We use the superspace
conventions of \cite{Leigh:1995ep}.} of the field $\Phi^i$. 
The coefficient $\mathcal{B}_1^{\prime} $ is given by
\begin{gather} \label{Bstrich}
\mathcal{B}_1^{\prime}=\frac{1}{16\pi^2}
\left(3C_2(G)- \sum\limits_{i=1}^{n} T(R_i)\right),
\end{gather}
where $T(R_i)$ is the Dynkin index in the representation $R_i$ of $G$,
$\mathrm{tr}(T^AT^B)_{R_i}=T(R_i)\delta^{AB}$. 

We now generalise (\ref{LS}) in two ways: We consider the off-shell case
and we couple the quantised theory to a classical supergravity
background as in section 2.4. Then the supercurrent anomaly is given by
\begin{gather} \label{scm}
\bar{\D}^{\da} \frac{\delta}{\delta H^{\al \da}} \Gamma' = \, \frac{1}{3} \, \D_\al \left[
\left( \phi \frac{ \delta}{\delta \phi} - \Phi^i \frac{\delta}{\delta \Phi^i}
 \right) \Gamma' \right] \, ,
\end{gather}
where the r.h.s.~is the transformation of the action under the super Weyl
transformation given by $\delta \phi = \sigma \phi$, $\delta \Phi^i = - \sigma
\Phi^i$ with $\sigma(z)$ the (chiral) Weyl transformation parameter.\footnote{For a
  detailed analysis of the transformation properties of quantum actions and
  of the quantum supercurrent under superconformal transformations, 
see~\cite{Erdmenger:1998tu,Erdmenger:1998xv,Erdmenger:1999uw}.} 
(\ref{scm}) generalises (\ref{scano}) to the matter case.
$\Gamma^\prime$ is the vertex functional obtained from the classical action
(\ref{ac}), together with the necessary counterterm
(\ref{counterterm2}) to guarantee $\Delta^-
\Gamma^\prime =0$ with $\Delta^-$ as in (\ref{plusminus}). 
As discussed in section 2.4, the supercurrent
in the local coupling theory on
curved superspace has a one-loop gauge anomaly. Therefore instead of the flat
space equation (\ref{LS}) we may now write
\begin{gather} \label{csp}
\left( \phi \frac{ \delta}{\delta \phi} - \Phi^i \frac{\delta}{\delta \Phi^i}
 \right) \Gamma' \, = \, - \,
\frac{{\mathcal B}_1{}'}{2} \phi^3  \,
{\rm tr} \, (W^\al W_\al) + \, \frac{1}{4} \phi^3 (\bar \D^2 +R) \, 
  \sum\limits_{i=1}^{n} \gamma_i 
  \bar \Phi_i e^{2V} \Phi^i \, ,
\end{gather}
which generalises (\ref{TTT}) and (\ref{wti}). 
The anomalous Weyl transformation (\ref{csp}) 
is the starting point for deriving the
local Callan-Symanzik equation in the presence of matter fields. 
It should be kept in mind that there are also contributions from the
gravitational anomalies to (\ref{csp}). These are discussed separately
in section \ref{external}.

Next we combine (\ref{csp}) with the Konishi anomaly. The Konishi anomaly
corresponds to an anomaly in an axial symmetry under which the matter fields
transform as $\delta \Phi^i = i v  \Phi^i$, $ v \in {\mathbb R}$ . 
The classical action
is invariant under this symmetry,
\begin{gather} \label{konishi} \left[
\int\! d^6z \, \Phi^i \frac{\delta}{\delta \Phi^i} \, - \,
\int\! d^6\bar z \, \bar \Phi^i \frac{\delta}{\delta \bar \Phi^i}  \right]
 \, S \, = \, 0 \, .
\end{gather}
The current associated with this symmetry 
is precisely the Konishi current: The local version of (\ref{konishi}) is
\begin{gather}
\Phi^i\frac{\delta}{\delta\Phi^i}S=
- \frac{1}{4} 
\phi^3 (\bar{\mathcal{D}}^2+R)(\bar{\Phi}_i e^{2V} \Phi^i) \, .
\end{gather}
In the quantised theory, the Konishi current has
an anomaly~\cite{Konishi:1983hf} which takes the off-shell form\footnote{again
in the conventions used in \cite{Leigh:1995ep} for the Konishi anomaly.}
\begin{gather}
\Phi^i\frac{\delta}{\delta\Phi^i}\Gamma^{\prime}
= \frac{1}{4} \phi^3 (\bar{\mathcal{D}}^2+R)(\bar{\Phi}_ie^{2V}\Phi^i)
 \, - \,
\frac{1}{16\pi^2}  \sum\limits_{i=1}^{n}
T(R_i)\, \phi^3 \mathrm{tr}(W^{\alpha}W_{\alpha}) \, ,
\end{gather}
or 
\begin{gather} \label{kon}
  \gamma_i 
\Phi^i\frac{\delta}{\delta\Phi^i}\Gamma^{\prime}
= \gamma_i \frac{1}{4} \phi^3 (\bar{\mathcal{D}}^2+R)(  
\bar{\Phi}_i e^{2V}\Phi^i) \, - \,
\frac{1}{16\pi^2}  \sum\limits_{i=1}^{n} \gamma_i 
T(R_i) \, \phi^3 \mathrm{tr}(W^{\alpha}W_{\alpha}) \, .
\end{gather}
Combining (\ref{kon}) and (\ref{csp}) we obtain the local Callan-Symanzik
equation 
\begin{gather} \left[
\phi \frac{\delta}{\delta \phi} - (1- \gamma_i)
\Phi^i\frac{\delta}{\delta\Phi^i} \right]  \Gamma^{\prime} 
\, = \, 
\left(\mathcal{B}'_1 + \frac{2}{16 \pi^2} \sum\limits_{i=1}^n 
\gamma_i T(R_i)\right) \frac{\delta}
{\delta \lambda } \Gamma^{\prime}  \, . 
\end{gather} 
Using 
\begin{gather} \label{cseqnPhi}
\mu\frac{\partial}{\partial\mu}\Gamma^{\prime}\, + \,   \left(\int\!
d^6z \,\left( \phi\frac{\delta}{\delta\phi} - \Phi^i\frac{\delta}{\delta\Phi^i} 
 \right) \, +   c.c. \right) \, \Gamma^{\prime}
\, = \, 0 \, ,
\end{gather}
which generalises (\ref{cseqn}), and the definition of $\Delta^+$ given in
(\ref{plusminus}), we have
\begin{equation} 
\left[\mu\frac{\partial}{\partial\mu} + 
\left(\B^{\prime}_1+ \frac{2}{16\pi^2}   \sum\limits_{i=1}^n \gamma_i \right)
\Delta^+ - \gamma_i {\cal N}_i
\right]\Gamma^{\prime}=0,
\, 
\end{equation}
with
\begin{equation}
{\cal N}_i \equiv \int\!
d^6z \,\left( \Phi^i\frac{\delta}{\delta\Phi^i}+c.c.\right) \, .
\end{equation}
By virtue of the shift equation (\ref{shifta}) we obtain the standard
Callan-Symanzik equation
\begin{gather}
\left[\mu\frac{\partial}{\partial\mu} + 
\left(\B^{\prime}_1+ \frac{2}{16\pi^2}   \sum\limits_{i=1}^n \gamma_i \right)
\left(\frac{-g^3}{1-\mathcal{A}_1g^2}\right)\partial_g  - \gamma_i {\cal N}_i
\right]\Gamma^{\prime}=0 \, .
\end{gather}
Thus with ${\cal B}'_1$ given by (\ref{Bstrich})
we find for the $\beta$-function that
\begin{gather}
\beta(g)=-\frac{g^3}{16\pi^2}\left(\frac{3C_2(G)-\sum_{i=1}^{n}(T(R_i)
-2\gamma_iT(R_i))}{1-C_2(G)g^2/8\pi^2}\right),
\end{gather}
which coincides with the NSVZ beta function.

It is straightforward to include a superpotential with local chiral
supercoupling into this discussion. In this case the matter part of the action
is given by

\begin{equation}\label{matteract2}
S_{\rm matter}
= \frac{1}{4} 
\int\! d^8z \, \tilde E \, 
\bar{\Phi}_i e^{2V} \Phi^i +\frac{1}{3!}\int\! d^6z \, \phi^3 Y_{ijk}
\Phi^i\Phi^j\Phi^k+\frac{1}{3!}\int\! d^6\bar{z} \, \bar \phi^3
\bar{Y}^{ijk}\bar{\Phi}_i\bar{\Phi}_j\bar{\Phi}_k.
\end{equation} 

In this theory the matter beta function satisfies
\begin{gather} 
\beta^{ijk}(Y)=3\gamma^{(i}_{\;m}Y^{jk)m}.\label{betamatt}
\end{gather}
As before the classical action is invariant under the Konishi symmetry, under
which the matter fields and the local matter couplings transform as
\begin{align} \label{kos}
&\Phi^i \rightarrow e^{i v}\Phi^i, \;\;\; Y_{ijk}\rightarrow
e^{-3i v}Y_{ijk},\\
&\bar{\Phi}_i\rightarrow e^{-i v}\bar{\Phi}_i,\; \bar{Y}^{ijk}
\rightarrow e^{3i v}\bar{Y}^{ijk} \, ,  \qquad v \in {\mathbb R} \, .
 \nonumber
\end{align}
The superpotential remains invariant under this transformation, as well as the
matter kinetic term. Note also that this symmetry differs from R
symmetry since it leaves the chiral compensator invariant.
The symmetry (\ref{kos})  leads to the Ward identity
\begin{gather}\label{konishi2}
\left(\int\! d^6z \, \left[3Y^{ ijk}\frac{\delta}{\delta
      Y^{ijk}}-\Phi^i\frac{\delta}{\delta\Phi^i}\right]-c.c.\right)S = 0 \, .
\end{gather}
In the quantised theory the Konishi anomaly gives rise to
\begin{gather} \label{kon3}
\left( \Phi^i\frac{\delta}{\delta\Phi^j} - 3 Y^{ijk} \frac{\delta}
{\delta Y^{ijk} } \right) \Gamma^{\prime}
= \frac{1}{4} \phi^3 (\bar{\mathcal{D}}^2+R)(\bar{\Phi}_i e^{2V}\Phi^i) \, - \,
\frac{1}{16\pi^2} \sum\limits_{i=1}^{n}T(R_i)
\phi^3\mathrm{tr}(W^{\alpha}W_{\alpha}) \, .
\end{gather}
Note that due to the variation with respect to the supercoupling on the
l.h.s., there is no contribution from the superpotential on the r.h.s. of 
(\ref{kon3}).  The absence of an anomaly involving the superpotential from this
equation may be seen as follows: As discussed for instance in 
\cite{Piguet:1986ug}, for renormalisable massless 
SUSY field theories there is
no superpotential contribution to the conformal Ward identity
(\ref{LS}). This is a consequence of R symmetry and 
leads in particular to  the relation 
(\ref{betamatt}) between the matter beta and gamma
functions. Therefore there is no superpotential contribution to the r.h.s.~of
(\ref{csp}). The same argument implies that there is no superpotential
contribution to the r.h.s.~of (\ref{kon3}).~\footnote{We are
  grateful to E.~Sokatchev for a discussion on this point. See \cite{Sokatchev}
for cases where mixing occurs.}

By adding (\ref{kon3}) to the integrated off-shell super Weyl identity
(\ref{csp}) we obtain the Callan-Symanzik equation
\begin{gather}
\left[\mu\frac{\partial}{\partial\mu} + 
\beta(g) \partial_g  + \left( \int\! d^6z \, \beta^{ijk} \frac{\delta}
{\delta Y^{ijk} } +  \int\! d^6\bar z \, \bar \beta_{ijk} \frac{\delta}
{\delta \bar Y_{ijk} }  \right) - \gamma_i {\cal N}_i
\right]\Gamma^{\prime}=0 \, . \label{ccs}
\end{gather}
with the same NSVZ gauge $\beta$-function as before. There are
additional contributions from the gravitational anomalies
to the r.h.s.~of  (\ref{ccs}) which we discuss next.

\section{External Anomalies -   Central Functions}\label{external}
\setcounter{equation}{0}

In addition to the internal anomalies discussed
above, there will also be external anomalies involving the super Euler density
and the square of the supergravity Weyl tensor. These external anomalies
appear both in the topological  and in the conformal
Ward identities. We calculate the coefficient of the gravitational anomalies
in the topological Ward identity (\ref{ponteq}) 
to one-loop order and show how an anomaly
shift similar to the one performed in section 2 above allows us to derive an
all-order expression for the central charge $c$ (the coefficient of the Weyl
tensor squared) in a particular well-defined renormalisation scheme. This
expression coincides with the one found in
\cite{Anselmi:1997am} based on a two-loop result of
Jack \cite{Jack:wd}.

\subsection{Gauge field contribution}

As before we first consider pure gauge theory.
In analogy to the gauge anomaly in the topological Ward identity we 
expect a gravitational  anomaly of the form
\begin{gather}\label{extanom1}
\Delta^{-}\Gamma^{\prime}=\frac{\mathcal{F} N_V}{24\pi^2} \, W^2 
-\frac{\mathcal{G} N_V}{24\pi^2} \, E^2,
\end{gather}
where
\begin{align}
W^2 =&\left(\int\! d^6z\,\phi^{3}\lambda^{\prime
    -1}(z)W^{\alpha\beta\gamma}W_{\alpha\beta\gamma}-\int\!
  d^6\bar{z}\, \bar{\phi}^{3}\bar{\lambda}^{\prime
    -1}(\bar{z})\bar{W}_{\dot{\alpha}\dot{\beta}\dot{\gamma}}
  \bar{W}^{\dot{\alpha}\dot{\beta}\dot{\gamma}} \right) \, , \nonumber \\
E^2 =& \int\! d^6z\, \phi^{3}\lambda^{\prime
-1}(z)(W^{\alpha\beta\gamma}W_{\alpha\beta\gamma}+(\mathcal{\bar{D}}^2+R)
[G^aG_a+2R\bar{R}])\nonumber \\
&-\int\! d^6\bar{z}\,\bar{\phi}^{3}\bar{\lambda}^{\prime -1}(\bar{z})
(\bar{W}_{\dot{\alpha}\dot{\beta}\dot{\gamma}} \bar{W}^{\dot{\alpha}
\dot{\beta}\dot{\gamma}}+(\mathcal{D}^2+\bar{R})[G^aG_a+2R\bar{R}]).
\end{align} 
Here $ W_{\alpha\beta\gamma} $ is the super-Weyl tensor and
$(W^{\alpha\beta\gamma}W_{\alpha\beta\gamma}+(\mathcal{\bar{D}}^2+R)[G^aG_a+2R\bar{R}])$
is the chiral projection of the the super Euler
density~\cite{Buchbinder:qv,Gates:nr}. 
By performing the one-loop calculation of appendix A.2,
we find for the coefficients ${\cal F}$, ${\cal G}$ to one-loop order:
\begin{gather}
\mathcal{F}_1 =  \, - \frac{1}{32}\, , \; \; 
\mathcal{G}_1 = \, - \frac{3}{16}\, . \label{FG}
\end{gather}
A summary of the one-loop calculation of appendix A.2 is as follows: 
We decompose (\ref{extanom1})
into components and perform the required triangle diagram computations
for the topological current, in analogy to the gauge anomaly computation
outlined in (\ref{Jtop}), (\ref{introcpnts}). In particular it is necessary to
calculate one-loop three point functions  which contain the the
topological current (\ref{Jtop}) and two copies of either the R current or the
energy-momentum tensor. - 
In the subsequent we assume that an appropriate scheme may be chosen
such that there are no higher order contributions to ${\cal F}$. This
gives a result for $c$ consistent with expectations. As far as ${\cal
G}$ is concerned, we work with its lowest-order value here and leave
an investigation of its higher order contributions to future
investigations. 

In analogy to the discussion of section 2.2,  the anomaly (\ref{extanom1}) 
may be shifted to the `shift 
equation' (\ref{qshift2}) 
by adding an appropriate counterterm to the action.
This counterterm reads
\begin{align}
\Gamma^{ct}(W,E)=&-\frac{\mathcal{F}_1 N_V}{24\pi^2} 
\int\! d^6z\, \phi^{3}[\ln(\lambda^{\prime}(z))+\ln 2 g^2]
W^{\alpha\beta\gamma}W_{\alpha\beta\gamma}\nonumber \\
&-\frac{\mathcal{F}_1 N_V}{24\pi^2} 
\int\! d^6\bar{z}\, \bar{\phi}^{3}[\ln(\bar{\lambda}^{\prime}(\bar{z}))+\ln
2 g^2]\bar{W}_{\dot{\alpha}\dot{\beta}\dot{\gamma}}
\bar{W}^{\dot{\alpha}\dot{\beta}\dot{\gamma}} \nonumber \\
& + \frac{\mathcal{G}_1N_V}{24\pi^2}\int\! d^6z \,\phi^{3}
\ln(\lambda^{\prime}(z))(W^{\alpha\beta\gamma}W_{\alpha\beta\gamma} 
+ (\mathcal{\bar{D}}^2+R)[G^aG_a+2R\bar{R}]) 
\nonumber \\
&+\frac{\mathcal{G}_1N_V}{24\pi^2}\int\!
d^6\bar{z}\, \bar{\phi}^{3}\ln(\bar{\lambda}^{\prime}(\bar{z}))
(\bar{W}_{\dot{\alpha}\dot{\beta}\dot{\gamma}}
\bar{W}^{\dot{\alpha}\dot{\beta}\dot{\gamma}} +
(\mathcal{D}^2 +\bar{R})[G^aG_a+2R\bar{R}]) .
\end{align}
The $\ln 2 g^2$ is necessary as before to obtain a well-defined constant
coupling limit. 
Note since the Euler density is topological, the terms in which it is
multiplied by the constant $\ln 2 g^2$ actually vanish, and such terms are
absent from the counterterm above.
For 
\begin{equation} \label{Gamma2strich}
\Gamma^{\prime \prime} = \Gamma' + \Gamma^{ct}(W,E)
\end{equation}
the quantum Pontryagin
equation is anomaly free,
\begin{gather}
\Delta^{-}\Gamma^{\prime\prime}= 0 \, ,
\end{gather}
but the shift equation (\ref{qshift2}) becomes anomalous as in the case of the
internal anomaly discussed in section 2. Let us derive the exact form
of this external anomaly.

For this purpose we first
 determine how the topological gravitational anomaly contributes to the
conformal anomaly. We begin by considering the contributions 
of the gravitational anomaly to the supercurrent divergence,
\begin{align}
\bar{\D}^{\dot{\alpha}}T_{\alpha\dot{\alpha}}=&  \,
\D_{\alpha}(T_{internal}
+T_{external}) \, , \\
T_{internal} =-&\frac{1}{6} \mathcal{B}_1 \, \phi^3W^{\beta}W_{\beta} \,
,\, \qquad  \mathcal{B}_1=\frac{3}{16\pi^2}C_2(G), \nonumber \\
T_{external}=&-\frac{c_1-a_1 }{48\pi^2}\, \phi^3
W^{\alpha\beta\gamma}W_{\alpha\beta\gamma} \nonumber \\
&+\frac{a_1}{48\pi^2}\, 
\phi^3 (\mathcal{\bar{D}}^2+R)(G^aG_a+2R\bar{R}) \label{T}
\end{align}
Just as the internal anomaly contributes with the one-loop coefficient
${\cal B}_1 $ to the superconformal anomaly (see (\ref{consistency2})
and (\ref{wti})), 
the gravitational anomaly
contributes with one-loop coefficients $c_1$ and $a_1$ given
by\footnote{For the pure gauge theory considered here, 
$N_\chi = 0$ but we include
the matter contribution for reference in the next section.} 
\begin{gather} \label{ca}
c_1 =  {\textstyle \frac{1}{24}} ( 3 N_V + N_\chi) \, , \quad
a_1 =  {\textstyle \frac{1}{24}} (9 N_V + N_\chi) \, .
\end{gather}
(\ref{T}) implies that
in the presence of the supergravity background, the local
Callan-Symanzik equation  (\ref{susylocalRG}) reads
\begin{align}
\phi\frac{\delta}{\delta\phi}\Gamma^{\prime\prime}=\, 
\mathcal{B}_1\frac{\delta}{\delta\lambda}\Gamma^{\prime\prime}
-&\frac{c_{1}-a_1}{16\pi^2} \, \phi^3 W^{\alpha\beta\gamma}
W_{\alpha\beta\gamma} \nonumber \\
+&\frac{a_1}{16\pi^2}\, \phi^3 (\mathcal{\bar{D}}^2+R)[G^aG_a+2R\bar{R}] \, .
\end{align}

The shift equation (\ref{shifta}) now becomes
\begin{gather}
\mathcal{B}_1\Delta^{+}\Gamma^{\prime\prime}=\beta(g)\partial_{g}\Gamma^{\prime\prime}+\frac{2\mathcal{F}_1N_V}{24\pi^2}\left(\frac{\beta(g)}{g}\right)\left(\int\! d^6z\,
\phi^{3}W^{\alpha\beta\gamma}W_{\alpha\beta\gamma}+c.c \right), \label{sb}
\end{gather} 
where we use the scheme, given by (\ref{shifta}),
in which $\beta(g)$ is the NSVZ $\beta$-function.
Note again the absence of the Euler density from the shift
equation (\ref{sb}), 
since this is a topological invariant and its integral vanishes.
The Callan-Symanzik equation now reads
\begin{gather}
\left(\mu\frac{\partial}{\partial
\mu}+\beta(g)\partial_g\right)\Gamma^{\prime\prime}=
\frac{1}{16\pi^2}\left(c_1- \frac{4}{3} \mathcal{F}_1 N_V 
\frac{\beta(g)}{g} \right) \left(\int\! d^6z \,
\phi^{3}W^{\alpha\beta\gamma}W_{\alpha\beta\gamma}+c.c \right). 
\end{gather}
From this expression we read off the result for the central charge 
$c$ to all orders,
\begin{gather}
c(g)=  c_1 - \frac{4}{3} N_V \mathcal{F}_1\frac{\beta(g)}{g} \, .
\end{gather}
Using the one-loop result (\ref{FG}) for
$\mathcal{F}_1$  we have
\begin{gather}
c(g)= c_1 + \frac{N_V}{24} \frac{\beta(g)}{g}.
\end{gather}
This coincides with the result of \cite{Anselmi:1997am}.

\subsection{Matter contribution}

Let us no include matter fields into the discussion of the section 3.1
in view of deriving an expression for the central charge $c$ in the
presence of matter. 
In generalisation of
 (\ref{csp}) we have, including the gravitational anomaly, 
\begin{align} \label{cspc}
\left( \phi \frac{ \delta}{\delta \phi} - \Phi^i \frac{\delta}{\delta \Phi^i}
 \right) \Gamma' \, = & \, - \,
\frac{{\mathcal B}_1{}'}{2} \phi^3  \,
{\rm tr} ( W^\al W_\al) + \, \frac{1}{4} \phi^3 (\bar \D^2 +R) \, 
  \sum\limits_{i=1}^{n} \gamma_i 
  \bar \Phi_i e^{2V} \Phi_i \, \nonumber\\ & - \frac{c_1 - a_1}{16 \pi^2}
 \phi^3 W^{\al \beta \gamma}W_{\al \beta \gamma} + \frac{a_1}{16 \pi^2}  \phi^3
 (\bar \D^2 +R) ( G^{\al \da}G_{\al \da} +2 R \bar R) \, ,
\end{align}
where $c_1$, $a_1$ are the one-loop coefficients (\ref{ca}) 
of the gravitational Weyl
anomaly. Similarly there are now gravitational anomalies contributing to the
Konishi identity as well,
\begin{align} \label{Konishicurved}
\Phi^i\frac{\delta}{\delta\Phi^i}\Gamma^{\prime}
= &
\frac{1}{4} \phi^3 (\bar{\mathcal{D}}^2+R)(\bar{\Phi}_ie^{2V}\Phi^i) \, - \,
\frac{1}{16\pi^2}\sum\limits_{i} 
T(R_i) \, \phi^3 \mathrm{tr}(W^{\alpha}W_{\alpha}) \nonumber\\
& - \frac{1}{24 \pi^2} (\H - \I) 
\phi^3 W^{\al \beta \gamma}W_{\al \beta \gamma} + \frac{\I}{24 \pi^2}  \phi^3
 (\bar \D^2 +R) ( G^{\al \da}G_{\al \da} +2 R \bar R) \, .
\end{align}
We calculate $\H$, $\I$ to one loop in appendix \ref{p calculation}  and find
\begin{gather}\label{Hres}
\H_1 = - \frac{1}{16} \, \sum\limits_{i} N_\chi{}^i  \, , \quad \I_1 = 0 \, .
\end{gather}
$N_\chi{}^i$ is the number of chiral fields with anomalous dimension
$\gamma_i$. 
For example, in SQCD with $N_f$
flavours, there is only one $\gamma$ and $N_\chi= 2 N_f N_c$.
The fact the $\I$ vanishes at one loop agrees with the fact that there is no
contribution linear in $\gamma$ to the central charge $a$, as discuss
in section 4 below. 

By combining the Weyl identity (\ref{cspc}) and the Konishi identity
(\ref{Konishicurved}) as before, and using $\Gamma^{\prime \prime}$ defined in
(\ref{Gamma2strich}), we obtain
\begin{align}
\left(\mu \frac{\partial}{\partial \mu} - \gamma_i \N_i \right) \Gamma^{\prime \prime}
= & - (\B_1' + \frac{2}{16\pi^2}   \sum\limits_{i=1}^n \gamma_i) \Delta^+
\Gamma^{\prime \prime} \nonumber\\ & + \frac{1}{16 \pi^2} \left( c_1 -
\frac{\gamma_i}{24} {N_\chi{}^i} \right) \left( \int\! d^6z \, \phi^3\, 
W^{\al \beta \gamma}W_{\al \beta \gamma} \; + c.c. \right) \, .
\end{align}
Therefore we have
\begin{gather} \label{finalCS}
\left(\mu \frac{\partial}{\partial \mu} + \beta(g) \partial_g 
- \gamma_i \N_i \right) \Gamma^{\prime \prime} = \, \frac{1}{16 \pi^2}
\, c \, \left( \int\! d^6z \, \phi^3\, 
W^{\al \beta \gamma}W_{\al \beta \gamma} \; + c.c. \right) \, ,
\end{gather}
with the central charge
\begin{gather}
c = c_1 + \frac{1}{24}\left(  N_V \frac{\beta_g}{g} - 
{\gamma_i} {N_\chi{}^i} \right) \, , \qquad c_1 = {
\frac{1}{24}} (3 N_V + \sum\limits_i
N_\chi{}^i) \, . \label{cresult}
\end{gather}
This all-order expression
coincides with the result of~\cite{Anselmi:1997am}.\footnote{
Note that the anomalous dimension used in \cite{Anselmi:1997am}
is twice the size of the one used here. At the same time $N_\chi=N_f
N_c$ in \cite{Anselmi:1997am}, whereas our conventions give $N_\chi=2 N_f
N_c$ for the theories considered in
\cite{Anselmi:1997am}.}

\section{Conclusions}\label{conclusion}
\setcounter{equation}{0}

In this paper we have given a derivation of the NSVZ beta function and of the
central charge $c$ in a well-determined renormalization scheme. Our derivation
is based on the topological anomaly which is present when the couplings are
allowed to be space-time dependent. 

As an outlook we comment on the implications of our results on the central
charge $a$, the coefficient of the Euler density anomaly which is expected to
be of relevance for generalizations of the C theorem to four dimensions. In
\cite{FreedmanOsborn}, a four-loop expression for a candidate
C function $\tilde a$ was obtained. Here we chose a particular renormalisation
scheme in which the result of \cite{FreedmanOsborn} reads
\begin{gather}
\tilde a \equiv a + \beta^i w_i \, , \quad \tilde a = a_1 - \frac{1}{8} {\rm \tr}
(\gamma \gamma) + \frac{1}{12} {\rm tr} (\gamma \gamma \gamma) + \frac{1}{4}
N_V \frac{\beta(g)}{g} + \frac{1}{3} Y^{ijk}\beta_{ijk} \label{FO}  \, .
\end{gather}
Here $w_i$ is a one-form in coupling space which may be identified with the
coefficient of a local conformal anomaly involving derivatives of the
couplings. For a derivation of (\ref{FO}) valid to all orders using the approach presented in this
paper, it would be necessary to calculate the coefficients of the new 
anomalies - for instance of ${\cal H}$ and ${\cal I}$ in \ref{Konishicurved} - 
to higher order, which we have not undertaken here.  So far our statements are
limited to expressions linear in the beta and gamma functions.
However the results of this paper are consistent with (\ref{FO}) in at least
two respects. First, the fact that ${\cal I} = 0$  at
one loop in (\ref{Konishicurved}) is
in agreement with the fact that there is no contribution of the form ${\rm tr}
(\gamma)$ (linear in $\gamma$ without further 
contributions from the couplings)  
to $\tilde a$ as given by (\ref{FO}):
A term of the form ${\cal I} {\rm tr}(\gamma)$ would enter $a$ when
adding (\ref{Konishicurved}) multiplied by $\gamma$ to (\ref{cspc}).   
Secondly, the contribution $\frac{1}{4} N_V \frac{\beta(g)}{g}$ to (\ref{FO})
is consistent with ${\cal G} = - \frac{3}{16}$ in (\ref{FG}), just as ${\cal F}
 = - \frac{1}{32}$  leads to the $\frac{1}{24} N_V \frac{\beta(g)}{g}$
 contribution to $c$ as given by (\ref{cresult}). 
The relative numerical factor is $4/3$  in both cases. However the derivation
of $c$ from ${\cal F}$ presented here does not apply to $a$ and ${\cal G}$
since the Euler density contribution is absent from the Callan-Symanzik
equation (\ref{finalCS}). - We intend to return to the question of
deriving an all-order expression for $a$ within the local coupling
approach in the future. 

Finally let us note that local couplings appear naturally within the AdS/CFT
correspondence and its generalisations since the couplings appear as boundary
values of the supergravity fields. In particular the holographic C theorem of
\cite{FGPW}, valid for field theories in arbitrary dimensions, may be
interpreted within field theory within the local coupling approach
\cite{Skenderis,Porrati,Erdmenger:2001ja}. We expect that the results
of the present paper will allow for a further understanding of the
field-theoretical implications of the holographic C theorem.

\bigskip \bigskip \bigskip\bigskip \bigskip \bigskip

{\bf Acknowledgements}

We are very grateful to Hugh Osborn for numerous discussions and helpful
comments. Moreover we thank E.~Sokatchev and B.~Wecht for useful discussions. 

The research of J.E.~is supported by DFG (Deutsche
Forschungsgemeinschaft) within the `Emmy Noether' programme, grant
ER301/1-4. J.B.~acknowledges support through a Research Fellowship of
the Alexander von Humboldt Foundation while in Berlin.

Part of the research of J.E. for this work was carried out at the KITP,  Santa
Barbara, USA.

\bigskip \bigskip \bigskip\bigskip \bigskip \bigskip

\appendix
\section{Appendix}
\setcounter{equation}{0}

\subsection{$\mathcal{N}=1$ SYM in components}

For completeness, 
we give here the super Yang-Mills action in components, which we use in the
calculations below. For compatibility with
\cite{Anselmi:1997am}, we use Euclidean signature in this appendix..
We have
\begin{gather}
S=S_D+  \, S_T \, 
\end{gather}
where $S_D$ is the dynamical part \cite{Erlich:1996mq}
\begin{gather}
S_D= \int\!
d^4x \, \frac{1}{2g^2}\mathrm{tr}[ F_{\mu\nu}F^{\mu\nu}+ 2
\bar{\lambda}\gamma^{\mu}D_{\mu}\lambda] \, ,
\end{gather}
and $S_T$ the topological part
\begin{gather}
S_T= \, - \frac{1}{2}\int\!
d^4x \, \tilde \theta \, 
\mathrm{tr}[F_{\mu\nu}^{\star}F^{\mu\nu}-2 \pr_{\mu}
(\bar{\lambda}\gamma^{\mu}\gamma_5\lambda)]\, , \qquad \tilde \theta =
\frac{\theta}{8 \pi^2} \, .
\end{gather}

\subsection{Calculating the coefficients of the gravitational
topological ano\-maly to
one loop}\label{k calculation}

To calculate the numerical values of $\mathcal{F}_1$ and $\mathcal{G}_1$, the
one-loop values of the coefficients defined in (\ref{extanom1}), we
use the results found
in~\cite{Erdmenger:1999xx,Anselmi:1997am,Erdmenger:1996yc} to evaluate 1-loop
triangle diagrams for the topological current corresponding to the Pontryagin
equation. In components, the supersymmetric expression $\Delta^{-}\Gamma'=0$
on flat space translates into (in Euclidean signature)

\begin{gather}
\frac{\delta}{\delta \tilde\theta} \Gamma' = \, - \, \nabla^\mu \langle J_\mu
  \rangle\, ,\\
J_{\mu} \equiv \,  \frac{1}{8} \left(
  \epsilon_{\mu\nu\rho\sigma}(A^{\nu a}
{\pr}^{\rho} A^{\sigma a } + \frac{1}{3}
A^{\sigma a} A^{\nu b} A^{\rho c} f_{abc})
-2\bar{\lambda}^a\gamma_{\mu}\gamma_5\lambda^a \right). \label{U}
\end{gather}
When coupled to the classical background fields $g_{\mu\nu},V_{\mu}$, the
above equation is anomalous such that
  
\begin{gather}\label{chern}
\frac{\delta}{\delta \tilde \theta} \Gamma' \, = \, - \, 
\nabla_{\mu}J^{\mu} +
\frac{g^2 k_1N_V}{24\pi^2}R_{\mu\nu\rho\sigma}\tilde{R}^{\mu\nu\rho\sigma}
+\frac{g^2 k_2N_V}{27\pi^2}V_{\mu\nu}\tilde{V}^{\mu\nu}.
\end{gather}
This is the local component  version of (\ref{extanom1}). 
First of all we note that although explicit factors of $g^2$ appear in the
gravitational anomaly in (\ref{chern}),
the anomaly coefficients $k_1$ and $k_2$ may be obtained to
one-loop order by a simple triangle diagram computation. This is due to the
unconventional normalization of the action (\ref{gauge}), which implies that
the R current and energy-momentum tensor involve a factor of $1/g^2$, whereas
every propagator has a factor of $g^2$. Thus, for instance in the one-loop 
triangle $\langle J RR\rangle$ with $J$ as in (\ref{U}) there is a resulting
overall factor of   $g^2$. 
Moreover, a further important point is that the factor of $27$ in the denominator of the second term is in agreement with
\cite{Intriligator:2003jj}. When comparing with the anomaly equation for the $R$
current derived in \cite{Anselmi:1997am},
\begin{gather}
\nabla_{\mu}R^{\mu}=\frac{c-a}{24\pi^2}R_{\mu\nu\rho\sigma}\tilde{R}^{\mu\nu\rho\sigma}+\frac{5a-3c}{9\pi^2}V_{\mu\nu}\tilde{V}^{\mu\nu},
\end{gather}
we see that the coefficients of the second anomaly term differ by a factor of
$1/3$. 
This takes account of the fact that there is no overall Bose
symmetrization in $\langle J RR \rangle$ as opposed to $\langle RRR \rangle$.
   
Decomposing (\ref{extanom1}) into components gives the following relation
between the coefficients in  (\ref{extanom1}) and in (\ref{chern}):
\begin{gather} \label{efge}
\mathcal{F}_1=\frac{1}{2}(5k_1+k_2), \qquad 
\mathcal{G}_1=\frac{1}{8}(3k_1+k_2).
\end{gather}
To determine $k_1$ and $k_2$, we calculate the
three point correlation functions $\langle U_{\mu}T_{\nu\rho}T_{\lambda\sigma}\rangle$ and
$\langle J_{\mu}R_{\nu}R_{\rho}\rangle$ to one-loop. 
To one loop order, 
the numerical values of $k_1$ and $k_2$ occurring in~(\ref{chern}) have
the following contributions:
\begin{gather}
k_1=\frac{1}{8}\left(4-2\left(\frac{-1}{2}\right)\right)\left(\frac{1}{4}\right)
=\frac{5}{32} \label{k1} \, .
\end{gather}
The factors present have the following origin: As shown in 
\cite{Erdmenger:1999xx}, \footnote{For the abelian case the gauge contribution 
to the first term in the anomaly given by (A.6) has been calculated in in
\cite{Dolgov}.}
and using the conventions of \cite{Anselmi:1997am}, the gauge field
contribution to the divergence of ${\langle JTT\rangle}$ is $1/2 \equiv 4/8$,
whereas the fermion contribution is $-1/8$. In the fermion contribution there
is a factor of $(-2)$ arising from the definition of $J$ in (\ref{U}) and a
factor of $1/2$ from the Majorana condition. Finally, there is an overall
factor of $1/4$ which is a remnant of the overall factor of $1/8$ in
(\ref{U}).   
On the other hand, the one-loop value of $k_2$ is obtained using the results
of \cite{Erdmenger:1996yc} or \cite{Erlich:1996mq} and is given by
\begin{gather}
k_2=(-2)\left(\frac{1}{2}\right)\left(\frac{1}{2}\right)\left(\frac{27}{8}\right)=-\left(\frac{27}{32}\right).
\end{gather}
The (-2) followed by the (1/2) arises from the definition of the J
current (\ref{U}), ,
whilst the next (1/2) is  the Majorana condition. The last factor is the
anomaly coefficient arising in the divergence of $\langle JRR \rangle$. 

These
values for $k_1$, $k_2$, when inserted into (\ref{efge}), 
 lead to the superspace coefficients given in
(\ref{FG}). 

\subsection{Calculating the gravitational anomaly of the Konishi
current to one loop}
\label{p calculation}

For calculating the gravitational anomalies contributing to
the divergence of the Konishi current, which in components, again using the
conventions of \cite{Anselmi:1997am},  reads
\begin{gather}
\nabla_\mu K^\mu= \frac{p_1 N_\chi}{24\pi^2}R_{\mu \nu \sigma \rho
}\tilde{R}^{\mu \nu
  \sigma \rho}+
\frac{p_2 N_\chi }{27\pi^2}V_{\mu \nu}\tilde{V}^{\mu \nu} \, ,
\end{gather}
with
\begin{gather}
K_\mu =  
\frac{1}{2} \bar \psi_i \gamma_\mu \gamma_5 \psi^i + \bar \phi_i 
\stackrel{\leftrightarrow }{D_{\mu }} \phi^i \, .
\end{gather}
By decomposing the relevant supergravity expressions in (\ref{Konishicurved})
we have
\begin{gather}
{\cal H} = N_\chi \, \frac{1}{2} (5 p_1 + p_2) \, , \qquad {\cal I} =  
N_\chi \, \frac{1}{2}(3p_1+ p_2) \, .
\end{gather}
This is obtained by multiplying (\ref{Konishicurved}) with $(\bar \D^2 +R)$, 
subtracting its complex conjugate, decomposing into components, restricting to
flat space and using the equations of motion. 

Calculating the relevant one-loop triangle diagrams, we find, 
using the
results of \cite{Erlich:1996mq,Erdmenger:1999xx} as before,
\begin{gather}
p_1 = - \frac{1}{16} \, , \qquad p_2 = \frac{3}{16} \, .
\end{gather}
This gives 
\begin{gather}
{\cal H}_1 = - \frac{1}{16} N_\chi \, , \qquad {\cal I}_1 = 0 
\,,
\end{gather}
which is used in (\ref{Hres}).  The fact that ${\cal I}$ = 0 at one loop
explains why no term linear in the anomalous dimension $\gamma$ contributes
to the coefficient of the Euler density $a$.



\bigskip


\newpage
\begin{thebibliography}{ll}

\bibitem{SV1}
M.~A.~Shifman and A.~I.~Vainshtein,
``On holomorphic dependence and infrared effects in supersymmetric gauge
theories,''
Nucl.\ Phys.\ B {\bf 359} (1991) 571.



\bibitem{LK}
L.~J.~Dixon, V.~Kaplunovsky and J.~Louis,
``Moduli Dependence Of String Loop Corrections To Gauge Coupling Constants,''
Nucl.\ Phys.\ B {\bf 355} (1991) 649.

\bibitem{Seiberg:1993vc}
N.~Seiberg,
``Naturalness versus supersymmetric nonrenormalization theorems,''
Phys.\ Lett.\ B {\bf 318} (1993) 469
[arXiv:hep-ph/9309335].

\bibitem{LK2}V.~Kaplunovsky and J.~Louis,
``Field dependent gauge couplings in locally supersymmetric effective quantum
field theories,''
Nucl.\ Phys.\ B {\bf 422} (1994) 57
[arXiv:hep-th/9402005].



\bibitem{Kraus:2001tg}
E.~Kraus,
``An anomalous breaking of supersymmetry in supersymmetric gauge 
theories  with local coupling,''
Nucl.\ Phys.\ B {\bf 620} (2002) 55
[arXiv:hep-th/0107239].


\bibitem{Kraus:2001id}
E.~Kraus,
 ``Calculating the anomalous supersymmetry breaking in super-Yang-Mills
theories with local coupling,''
Phys.\ Rev.\ D {\bf 65} (2002) 105003
[arXiv:hep-ph/0110323].

\bibitem{Kraus:2002nu}
E.~Kraus, C.~Rupp and K.~Sibold,
 ``Supersymmetric Yang-Mills theories with local coupling: The supersymmetric
gauge,''
Nucl.\ Phys.\ B {\bf 661} (2003) 83
[arXiv:hep-th/0212064].

\bibitem{KrausWZ}
E.~Kraus, C.~Rupp and K.~Sibold,
``Supercurrent and local coupling in the Wess-Zumino model,''
Eur.\ Phys.\ J.\ C {\bf 24} (2002) 631
[arXiv:hep-th/0205013].

\bibitem{Kraus:2002se}
E.~Kraus,
``Anomalies in quantum field theory: Properties and characterization,''
arXiv:hep-th/0211084.

\bibitem{Kraus:2001kn}
E.~Kraus and D.~St\"ockinger,
 ``Nonrenormalization theorems of supersymmetric QED in the Wess-Zumino
gauge,''
Nucl.\ Phys.\ B {\bf 626} (2002) 73
[arXiv:hep-th/0105028].

\bibitem{BPHZ}
O.~Piguet and S.~Sorella, `Algebraic Renormalization'', Springer
Verlag, Berlin 1995; \newline
J.~Collins, `Renormalization'', Cambridge University Press 1984.

\bibitem{Bos}
M.~Bos,
``Explicit calculation of the renormalized singlet axial anomaly,''
Nucl.\ Phys.\ B {\bf 404} (1993) 215
[arXiv:hep-ph/9211319].


\bibitem{Arkani-Hamed:1997mj}
N.~Arkani-Hamed and H.~Murayama,
``Holomorphy, rescaling anomalies and exact beta functions in  supersymmetric
gauge theories,''
JHEP {\bf 0006} (2000) 030
[arXiv:hep-th/9707133].

\bibitem{Novikov:ic}
V.~A.~Novikov, M.~A.~Shifman, A.~I.~Vainshtein and V.~I.~Zakharov,
``Supersymmetric Instanton Calculus: Gauge Theories With Matter,''
Nucl.\ Phys.\ B {\bf 260} (1985) 157
[Yad.\ Fiz.\  {\bf 42} (1985) 1499].

\bibitem{Osborn:gm}
H.~Osborn,
 ``Weyl Consistency Conditions And A Local Renormalization Group Equation For
General Renormalizable Field Theories,''
Nucl.\ Phys.\ B {\bf 363} (1991) 486.







\bibitem{Kraus:1992ru}
E.~Kraus and K.~Sibold,
``Local couplings, double insertions and the Weyl consistency condition,''
Nucl.\ Phys.\ B {\bf 398} (1993) 125.


\bibitem{Jack}
I.~Jack and H.~Osborn,
``Analogs For The C Theorem For Four-Dimensional Renormalizable Field Theories,''
Nucl.\ Phys.\ B {\bf 343} (1990) 647.


\bibitem{Zamolodchikov}
A.~B.~Zamolodchikov,
``'Irreversibility' Of The Flux Of The Renormalization Group In A 2-D Field Theory,''
JETP Lett.\ {\bf 43} (1986) 730.

\bibitem{Cardy:cw}
J.~L.~Cardy,
``Is There A C Theorem In Four-Dimensions?,''
Phys.\ Lett.\ B {\bf 215} (1988) 749.

\bibitem{Osborn:2003vk}
H.~Osborn,
``Local couplings and Sl(2,R) invariance for gauge theories at one loop,''
Phys.\ Lett.\ B {\bf 561} (2003) 174
[arXiv:hep-th/0302119].

\bibitem{FreedmanOsborn} 
D.~Z.~Freedman and H.~Osborn,
``Constructing a c-function for SUSY gauge theories,''
Phys.\ Lett.\ B {\bf 432} (1998) 353
[arXiv:hep-th/9804101].

\bibitem{Anselmi:1997am}
D.~Anselmi, D.~Z.~Freedman, M.~T.~Grisaru and A.~A.~Johansen,
 ``Nonperturbative formulas for central functions of supersymmetric gauge
theories,''
Nucl.\ Phys.\ B {\bf 526} (1998) 543
[arXiv:hep-th/9708042].


\bibitem{Intriligator:2003jj}
K.~Intriligator and B.~Wecht,
``The exact superconformal R-symmetry maximizes a,''
Nucl.\ Phys.\ B {\bf 667} (2003) 183
[arXiv:hep-th/0304128].

\bibitem{Intriligator:2003mi}
K.~Intriligator and B.~Wecht,
``RG fixed points and flows in SQCD with adjoints,''
Nucl.\ Phys.\ B {\bf 677} (2004) 223
[arXiv:hep-th/0309201].

\bibitem{Barnes:2004jj}
E.~Barnes, K.~Intriligator, B.~Wecht and J.~Wright,
``Evidence for the strongest version of the 4d a-theorem, via a-maximization
Nucl.\ Phys.\ B {\bf 702} (2004) 131
[arXiv:hep-th/0408156].

\bibitem{Barnes:2005zn}
E.~Barnes, K.~Intriligator, B.~Wecht and J.~Wright,
``N=1 RG Flows, Product Groups, and a-Maximization,''
arXiv:hep-th/0502049.


\bibitem{Kutasov:2003ux}
D.~Kutasov,
``New results on the 'a-theorem' in four dimensional supersymmetric field
theory,''
arXiv:hep-th/0312098.

\bibitem{Kutasov:2004xu}
D.~Kutasov and A.~Schwimmer,
``Lagrange multipliers and couplings in supersymmetric field theory,''
Nucl.\ Phys.\ B {\bf 702} (2004) 369
[arXiv:hep-th/0409029].


\bibitem{Buchbinder:qv}
I.~L.~Buchbinder and S.~M.~Kuzenko,
 ``Ideas And Methods Of Supersymmetry And Supergravity Or : A Walk Through
Superspace,'' 
Bristol, UK: IOP (1998), 656 p. 

\bibitem{Gates:nr}
S.~J.~Gates, M.~T.~Grisaru, M.~Rocek and W.~Siegel,
``Superspace, Or One Thousand And One Lessons In Supersymmetry,''
Front.\ Phys.\  {\bf 58} (1983) 1
[arXiv:hep-th/0108200].



\bibitem{Erdmenger:1998tu}
J.~Erdmenger, C.~Rupp and K.~Sibold,
``Conformal transformation properties of the supercurrent in four  dimensional
supersymmetric theories,''
Nucl.\ Phys.\ B {\bf 530} (1998) 501
[arXiv:hep-th/9804053].

\bibitem{Erdmenger:1998xv}
J.~Erdmenger and C.~Rupp,
``Superconformal Ward identities for Green functions with multiple
supercurrent insertions,''
Annals Phys.\  {\bf 276} (1999) 152
[arXiv:hep-th/9811209].


\bibitem{Erdmenger:1999uw}
J.~Erdmenger, C.~Rupp and K.~Sibold,
``Superconformal transformation properties of the supercurrent. II:  Abelian
gauge theories,''
Nucl.\ Phys.\ B {\bf 565} (2000) 363
[arXiv:hep-th/9907169].




\bibitem{Clark:1980dw}
T.~E.~Clark, O.~Piguet and K.~Sibold,
``The Renormalized Supercurrents In Supersymmetric QED,''
Nucl.\ Phys.\ B {\bf 172} (1980) 201.


\bibitem{Shifman:1986zi}
M.~A.~Shifman and A.~I.~Vainshtein,
``Solution Of The Anomaly Puzzle In Susy Gauge Theories And The Wilson Operator
Expansion,''
Nucl.\ Phys.\ B {\bf 277}, 456 (1986)
[Sov.\ Phys.\ JETP {\bf 64}, 428 (1986\ ZETFA,91,723-744.1986)].

\bibitem{Kogan:1995mr}
I.~I.~Kogan, M.~A.~Shifman and A.~I.~Vainshtein,
 ``Matching conditions and duality in N=1 SUSY gauge theories in the conformal
window,''
Phys.\ Rev.\ D {\bf 53} (1996) 4526
[Erratum-ibid.\ D {\bf 59} (1999) 109903]
[arXiv:hep-th/9507170].

\bibitem{Leigh:1995ep}
R.~G.~Leigh and M.~J.~Strassler,
``Exactly marginal operators and duality in four-dimensional N=1
supersymmetric gauge theory,''
Nucl.\ Phys.\ B {\bf 447} (1995) 95
[arXiv:hep-th/9503121].



\bibitem{Anselmi:1996dd}
D.~Anselmi, D.~Z.~Freedman, M.~T.~Grisaru and A.~A.~Johansen,
``Universality of the operator product expansions of SCFT(4),''
Phys.\ Lett.\ B {\bf 394} (1997) 329
[arXiv:hep-th/9608125].

\bibitem{Jack:wd}
I.~Jack,
 ``Background Field Calculations In Curved Space-Time. 3. Application To A
General Gauge Theory Coupled To Fermions And Scalars,''
Nucl.\ Phys.\ B {\bf 253} (1985) 323.


\bibitem{supercurrentPS}O.~Piguet and K.~Sibold,
``The Supercurrent In N=1 Supersymmetrical Yang-Mills Theories'',
Nucl.\ Phys.\ B {\bf 196} (1982) 428, 447.

\bibitem{supercurrentGZ}
M.~T.~Grisaru, B.~Milewski and D.~Zanon,
``Supercurrents, Anomalies And The Adler-Bardeen Theorem,''
Phys.\ Lett.\ B {\bf 157} (1985) 174;
\newline M.~T.~Grisaru, B.~Milewski and D.~Zanon,
``The Supercurrent And The Adler-Bardeen Theorem,''
Nucl.\ Phys.\ B {\bf 266} (1986) 589.



\bibitem{Piguet:1986ug}
O.~Piguet and K.~Sibold,
``Renormalized Supersymmetry. The Perturbation Theory Of N=1 Supersymmetric
Theories In Flat Space-Time,'' Birkh\"auser, Boston 1986, 346 p.



\bibitem{Konishi:1983hf}
K.~Konishi,
 ``Anomalous Supersymmetry Transformation Of Some Composite Operators In
Sqcd,''
Phys.\ Lett.\ B {\bf 135} (1984) 439.

\bibitem{Sokatchev}
B.~Eden, C.~Jarczak, E.~Sokatchev and Y.~S.~Stanev,
``Operator mixing in N = 4 SYM: The Konishi anomaly revisited,''
arXiv:hep-th/0501077.

\bibitem{FGPW}
D.~Z.~Freedman, S.~S.~Gubser, K.~Pilch and N.~P.~Warner,
``Renormalization group flows from holography supersymmetry and a  c-theorem,''
Adv.\ Theor.\ Math.\ Phys.\  {\bf 3} (1999) 363
[arXiv:hep-th/9904017].




\bibitem{Skenderis}
S.~de Haro, S.~N.~Solodukhin and K.~Skenderis,
``Holographic reconstruction of spacetime and renormalization in the  AdS/CFT
correspondence,''
Commun.\ Math.\ Phys.\  {\bf 217} (2001) 595
[arXiv:hep-th/0002230].



\bibitem{Porrati}
D.~Anselmi, L.~Girardello, M.~Porrati and A.~Zaffaroni,
``A note on the holographic beta and c functions,''
Phys.\ Lett.\ B {\bf 481} (2000) 346
[arXiv:hep-th/0002066].



\bibitem{Erdmenger:2001ja}
J.~Erdmenger,
 ``A field-theoretical interpretation of the holographic renormalization
group,''
Phys.\ Rev.\ D {\bf 64} (2001) 085012
[arXiv:hep-th/0103219].

\bibitem{Erlich:1996mq}
J.~Erlich and D.~Z.~Freedman,
``Conformal symmetry and the chiral anomaly,''
Phys.\ Rev.\ D {\bf 55} (1997) 6522
[arXiv:hep-th/9611133].

\bibitem{Dolgov}
A.~D.~Dolgov, I.~B.~Khriplovich, A.~I.~Vainshtein and V.~I.~Zakharov,
``Photonic Chiral Current And Its Anomaly In A Gravitational Field,''
Nucl.\ Phys.\ B {\bf 315} (1989) 138.



\bibitem{Erdmenger:1999xx}
J.~Erdmenger,
 ``Gravitational axial anomaly for four dimensional conformal field
theories,''
Nucl.\ Phys.\ B {\bf 562} (1999) 315
[arXiv:hep-th/9905176].



\bibitem{Erdmenger:1996yc}
J.~Erdmenger and H.~Osborn,
 ``Conserved currents and the energy-momentum tensor in conformally  invariant
theories for general dimensions,''
Nucl.\ Phys.\ B {\bf 483} (1997) 431
[arXiv:hep-th/9605009].









\end{thebibliography}
\end{document}